\documentclass{pic2012} %%% pdflatex pic2012_durr.tex

\def\runauthor{S.\,D\"urr}
\def\shorttitle{Recent Progress in Lattice QCD}

\usepackage{hyperref} %%% note: automatically loads "color", use "dvips -z"

\begin{document}

\title{RECENT PROGRESS IN LATTICE QCD}

\author{Stephan~D\"urr}

\address{Bergische Universit\"at Wuppertal,
Gau{\ss}stra{\ss}e\,20, 42119 Wuppertal, Germany\\
J\"ulich Supercomputing Center,
Forschungszentrum J\"ulich, 52425 J\"ulich, Germany\\
E-mail: durr\,(AT)\,itp.unibe.ch}

\maketitle

\abstracts{Recent progress in Lattice QCD is highlighted. After a brief
introduction to the methodology of lattice computations the presentation
focuses on three main topics: Hadron Spectroscopy, Hadron Structure and Lattice
Flavor Physics. In each case a summary of recent computations of selected
quantities is provided.}

%%%%%%%%%%%%%%%%%%%%%%%%%%%%%%%%%%%%%%%%%%%%%%%%%%%%%%%%%%%%%%%%%%%%%%%%%%%%%%%

\newcommand{\al}{\alpha}
\newcommand{\be}{\beta}
\newcommand{\ga}{\gamma}
\newcommand{\de}{\delta}
\newcommand{\ep}{\epsilon}
\newcommand{\ve}{\varepsilon}
\newcommand{\ze}{\zeta}
\newcommand{\et}{\eta}
\renewcommand{\th}{\theta}
\newcommand{\vt}{\vartheta}
\newcommand{\io}{\iota}
\newcommand{\ka}{\kappa}
\newcommand{\la}{\lambda}
\newcommand{\rh}{\rho}
\newcommand{\vr}{\varrho}
\newcommand{\si}{\sigma}
\newcommand{\ta}{\tau}
\newcommand{\ph}{\phi}
\newcommand{\vp}{\varphi}
\newcommand{\ch}{\chi}
\newcommand{\ps}{\psi}
\newcommand{\om}{\omega}

\newcommand{\Mpi}{M_\pi}
\newcommand{\Fpi}{F_\pi}
\newcommand{\Mka}{M_K}
\newcommand{\Fka}{F_K}
\newcommand{\Met}{M_\et}
\newcommand{\Fet}{F_\et}
\newcommand{\Mss}{M_{s\bar{s}}}
\newcommand{\Fss}{F_{s\bar{s}}}
\newcommand{\Mcc}{M_{c\bar{c}}}
\newcommand{\Fcc}{F_{c\bar{c}}}

\newcommand{\pad}{\partial}
\newcommand{\psl}{\partial\!\!\!/}
\newcommand{\til}{\tilde}
\newcommand{\pri}{^\prime}
\renewcommand{\dag}{^\dagger}
\newcommand{\<}{\langle}
\renewcommand{\>}{\rangle}
\newcommand{\gaf}{\gamma_5}
\newcommand{\nab}{\nabla\!}
\newcommand{\lap}{\triangle}
\newcommand{\dal}{{\sqcap\!\!\!\!\sqcup}}
\newcommand{\trc}{\mr{tr}}

\newcommand{\bdm}{\begin{displaymath}}
\newcommand{\edm}{\end{displaymath}}
\newcommand{\bea}{\begin{eqnarray}}
\newcommand{\eea}{\end{eqnarray}}
\newcommand{\beq}{\begin{equation}}
\newcommand{\eeq}{\end{equation}}

\newcommand{\mr}{\mathrm}
\newcommand{\mb}{\mathbf}
\newcommand{\ri}{\mathrm{i}}
\newcommand{\cd}{\!\cdot\!}
\newcommand{\Nf}{N_{\!f}}
\newcommand{\Nt}{N_{\!t}}
\newcommand{\Nc}{N_{\!c}}
\newcommand{\Nv}{N_{\!v}}
\newcommand{\Ns}{N_{\!s}}
\newcommand{\MeV}{\,\mr{MeV}}
\newcommand{\GeV}{\,\mr{GeV}}
\newcommand{\TeV}{\,\mr{TeV}}
\newcommand{\fm}{\,\mr{fm}}
\newcommand{\const}{\mr{const}}
\newcommand{\MSbar}{{\overline{\mr{MS}}}}

%%%%%%%%%%%%%%%%%%%%%%%%%%%%%%%%%%%%%%%%%%%%%%%%%%%%%%%%%%%%%%%%%%%%%%%%%%%%%%%

\section{Introduction \label{sec:introduction}}

%%%%%%%%%%%%%%%%%%%%%%%%%%%%%%%%%%%%%%%%%%%%%%%%%%%%%%%%%%%%%%%%%%%%%%%%%%%%%%%

Lattice QCD has come of age.
What is meant with this statement is that 40 years after its inception
\cite{Wilson:1974sk} this framework is now able to deliver clear-cut
predictions for a number of phenomenologically relevant quantities.

From an experimentalist's viewpoint what matters is the precision/accuracy
of a computation; to be relevant it must match the experiment's precision
(which often is a challenge).
From a theoretician's viewpoint the key issue is that all quantities are
well defined and that the total uncertainty of the final result (usually split
into statistical and systematic contributions) can be reliably assessed.
In that respect the last decade has brought a major improvement: The
quenched approximation (in which the quark loops that come from the functional
determinant are omitted) is gone.
While often found to have a small impact in practice (a posteriori, i.e.\ by
comparing to the full/unquenched result), this approximation was a nuisance,
because its impact was extremely hard to quantify.
Todays lattice computations set new standards, since the collaborations spend
an enormous amount of effort to control and quantify \emph{all} sources of
systematic uncertainty.

There is a number of physics questions in which Lattice QCD plays a key role.
One example is ``Do we understand the weight of this world'' (the visible
matter in the Universe) ?
With the Higgs particle established at the 4$\si$ level (as reviewed at this
conference \cite{ArthurSchaffer,RobertoSalerno}) one might think that
this is the cause.
Looking at actual numbers we realize that the Higgs mechanism is almost
irrelevant for the mass of the proton; the average of the up and down
quark masses is
$m_{ud}^\mr{phys}\equiv(m_u^\mr{phys}\!+\!m_d^\mr{phys})/2\simeq3.5\MeV$ in
$(\MSbar,2\GeV)$ conventions, while the proton mass is $M_p\simeq940\MeV$.
At this point we stand accused of comparing apples to pears; the former mass
is a scheme and scale dependent quantity, while the latter one is a truly
physical quantity.
A more meaningful attempt would relate the proton mass to the mass that the
nucleon would have in the 2-flavor chiral limit, that is to $M_N\simeq880\MeV$
with $m_{ud}\to0$ and all other quark masses held fixed
(cf.\ Sec.\,\ref{sec:structure}).
Still, the message is the same: The weight of the world which we experience
is essentially due to the relativistic dynamics that is involved in the process
by which the strong force binds quarks and gluons into protons (and other
color neutral objects).

To date the lattice is the only known theoretical tool which can ``solve QCD''
in practical terms and with fully controlled systematics.
This means that it is capable of establishing the link between fundamental
parameters of the Standard Model (e.g.\ quark masses or CKM matrix elements)
and the quantities measured in experiment (e.g.\ hadron masses or leptonic and
semi-leptonic widths or branching fractions).
In the following the goal will be to explain the theoretical underpinning of
these computations, and to highlight some of the latest results.

The remainder of this review is organized as follows.
Section~\ref{sec:lattice} gives an outline of the field-theoretic setup of
Lattice QCD computations.
The next three sections contain the core of the presentation, an update on
Hadron Spectroscopy (Sec.\,\ref{sec:spectroscopy}\,), Hadron Structure
(Sec.\,\ref{sec:structure}\,) and Flavor Physics (Sec.\,\ref{sec:flavor}\,,
with a brief presentation of the recent FLAG activities).
To give the reader an idea of what a tiny segment of lattice studies has been
covered, we shall attempt to list some of the topics omitted in
Section~\ref{sec:other} before presenting a summary in
Section~\ref{sec:summary}\,.

%%%%%%%%%%%%%%%%%%%%%%%%%%%%%%%%%%%%%%%%%%%%%%%%%%%%%%%%%%%%%%%%%%%%%%%%%%%%%%%

\section{Lattice Field Theory \label{sec:lattice}}

%%%%%%%%%%%%%%%%%%%%%%%%%%%%%%%%%%%%%%%%%%%%%%%%%%%%%%%%%%%%%%%%%%%%%%%%%%%%%%%

\subsection{Regulating QCD in the UV and the IR}

Quantum Chromodynamics (QCD) is a field theory.
As such it requires, in an intermediate stage, regularization both in the
ultraviolet (UV) and in the infrared (IR) to be mathematically well behaved.

The lattice approach (more precisely: the multitude of lattice approaches,
see below) does this by replacing the spacetime manifold by a regularly spaced
4-dimensional grid.
Typically, a hypercubic set of grid points $x\in(n_1,n_2,n_3,n_4)a$ is used,
with $n_i$ ($i=1,2,3$) running from 1 to $N_s$ and $n_4$ running from 1 to
$N_t$ and $a$ the lattice spacing.
The temporal direction is considered the fourth direction, since we work in
Euclidean space, where all directions have the same sign in the metric.
The lattice spacing $a$ is usually chosen isotropically, i.e.\ the same (with
$c=1$) in Euclidean space and time directions (though other options are
possible).

The matter and gauge fields live in these grid points and in the links that
connect neighboring grid points, respectively.
Either type of fields may have internal degrees of freedom (as indicated by
spinor and/or color indices).
For convenience one usually introduces periodic or antiperiodic boundary
conditions.
Hence, starting in $(aN_s,y,z,t)$ and hopping one unit in the 1-direction
one ends up in $(a\,1,y,z,t)$ and similarly for the other directions, except
for the 4-direction, where one picks up a sign if one is a fermion.
With this setup the total number of degrees of freedom is finite, while the
degrees of freedom themselves (the quark and gluon fields) are continuous
(if they are discrete, too, one has a spin system).

By introducing the rule that each configuration of quark and gluon fields shall
be attributed the Boltzmann weight $\exp(-S_\mr{QCD})$, where the action
\cite{Fritzsch:1973pi} %%% GattringerLang_(2.17)
\beq
S_\mr{QCD}=
\frac{1}{2g^2}\mr{Tr}(F_{\mu\nu}F_{\mu\nu})
+\sum_{i=1}^{\Nf}\bar q^{(i)}(D\!\!\!\!/+\!m^{(i)})q^{(i)}
+\ri\th\frac{1}{32\pi^2}\ep_{\mu\nu\rh\si}\mr{Tr}(F_{\mu\nu}F_{\rh\si})
\label{def_QCD}
\eeq
is the Euclidean counterpart of the Minkowskian Lagrangian, one defines the
partition function of QCD, provided there is an unambiguous measure for the
variables in each grid point or link [which is true for compact gauge groups
like $SU(\Nc)$].
This weight depends on the discretization or action chosen, i.e.\ on the
details of how one expresses the quantities in (\ref{def_QCD}) in terms of
the quark fields $q^{(i)}(x)$ and gauge links $U_\mu(x)$, where $x$ now denotes
a 4-dimensional coordinate [which is implicit in (\ref{def_QCD})].
For instance, the Dirac operator $D\!\!\!\!/\,$ may be anti-hermitian (as one
would expect from a continuum viewpoint) in one discretization, but not even a
normal operator in another one.
Phenomenologically, the parameter $\th$ is known to be extremely close to 0 (or
$\pi$), this is why the last term in (\ref{def_QCD}) is usually omitted.

To cut a long story short, the discretization of a field theory is an intermediate
step which renders the path-integral (or partition function) finite.
This happens right from the outset, in sharp contrast to perturbative
approaches where one tames, at a much later stage, integrations which otherwise
would produce infinities.
The lattice provides a UV cut-off (through $a>0$) and an IR-cut-off (through
$V=L^3T<\infty$, where $L=N_sa$ is the box-length and $T=N_ta$ the time-extent)
that respects the gauge invariance of (\ref{def_QCD}).
One then computes inverse correlation lengths in lattice units, e.g.\
$a/\xi_\pi=a\Mpi$ and $a/\xi_\Omega=aM_\Omega$.
The rule is that one must form ratios of such quantities before one is allowed
to take the continuum limit.
Hence, wile $\xi_\pi/a$ would diverge, the ratio $\xi_\pi/\xi_\Omega$ stays
finite under $a\to0$.
The crucial point is that the result is independent of the details of the
lattice action (\ref{def_QCD}), provided some guidelines were observed.
In other words, the various lattice regularizations provide different ways
of defining QCD, but they share a universal continuum limit.
Any valid discretization of (\ref{def_QCD}) is thus part of the
\emph{definition} of QCD !

\subsection{Scale hierarchies in Lattice QCD}

With this bit of lattice ideology in place, it is time to reflect on something
relevant in every days use of Lattice QCD, namely the scale hierarchies
involved.
Evidently, the shortest length-scale at hand is the lattice spacing $a$, the
largest one is the box size $L$.
The correlation lengths of the quarks that occur in (\ref{def_QCD}) and of the
hadrons that emerge as bound states should somehow lie in between these
extremes.

It turns out that for the shortest scales it is the (heaviest) quark mass that
matters, while for the largest scales it is the mass of the (lightest)
asymptotic state that is relevant.
In other words, for the quarks we need to pay attention that they do not
``fall through the grid'' (i.e.\ we request $am_q\ll1$), while for the pions
we need to make sure that they are not squeezed too much (i.e.\ we request
$\Mpi L\gg1$).
This may be a bit of a challenge, because maintaining both conditions
simultaneously requires having a large number of grid points in each direction.

At the time of writing the largest number of grid points that can be sustained
in current state-of-the-art lattice QCD simulations is of the order $128^4$.
Even with this somewhat optimistic assumption on $L/a$ the goal $\Mpi L>4$
would translate, at the physical pion mass, into $L=4.1/(135\MeV)=6\fm$ and
hence into $a=0.047\fm$.
With $m_c\simeq1.1\GeV$ the product is then $am_c=0.26$.
This number is small enough, but in order to take a continuum limit we
need to have several lattice spacings.
We can only afford to go to larger $a$; upon doubling the lattice spacing we
end up with $am_c=0.52$ which (with the actions currently in use) is barely
acceptable.

Hence simulating almost physical light quarks and, at the same time, charm
quarks without terrific cut-off effects poses enormous computational
requirements.
Usually one chooses for the up and down quarks a common mass $m_{ud}$, and
either this mass is kept in the vicinity of its physical (isospin averaged)
value or $m_c$ is kept near its physical mass.
Often some extrapolation in $m_{ud}$ and/or $m_c$ is performed, while $m_s$ is
usually simulated next to its physical mass value.
Simulating bottom quarks in the vicinity of their physical mass is out of
question (except for special setups); they are usually treated in entirely
different frameworks.
And the top is omitted from Lattice QCD calculations, since it is too short
lived to hadronize.

\subsection{How to remove systematic effects}

As emphasized in subsec.\,2.1, the lattice (or something equivalent) is a
necessary intermediate step in the definition of QCD.
In consequence, a computer code which yields a stochastic estimate
\cite{Creutz:1980zw} of the path integral (\ref{def_QCD}) for a given lattice
spacing $a$ and box size $L$, at a given set of quark masses $m_q=(m_{ud},m_s)$
[and possibly $m_c$], does not complete the job.
A good lattice calculation combines the information from several simulations to
remove the remnants of the lattice formulation.
Thus, when reading a lattice paper one should have the following questions in
mind:
\begin{description}
\itemsep-2pt
\item[(1)]
Has the continuum limit ($a\!\to\!0$) been taken ?
\item[(2)]
Are the finite-volume effects (from $L\!<\!\infty$) under control ?
\item[(3)]
Are the simulations performed anywhere close to $\Mpi\!=\!135\MeV$ ?
\item[(4)]
Advanced: are theoretical uncertainties properly assessed/propagated ?
\item[(5)]
Experts: algorithm details, treatment of isospin breakings, resonances, ...
\end{description}
Unfortunately, all the interesting directions tend to be expensive in terms of
computer time.
The computational requirements tend to increase roughly like
\beq
\mr{CPU}\propto 1/a^{4-6}\;,\qquad
\mr{CPU}\propto L^5\,,\qquad
\mr{CPU}\propto 1/m_q^{1-2}
\eeq
and the meaning is that in each case the other parameters are held fixed.
In reality, when pushing $m_{ud}$ down towards $m_{ud}^\mr{phys}\simeq3.5\MeV$
also the box length $L$ needs to be increased to maintain the bound $\Mpi L>4$.
In addition, there are algorithmic issues which lead to a proliferation of
noise near the chiral limit \cite{Schaefer:2012tq}; this makes simulations
close to the physical mass point even more demanding.

The quark fields in (\ref{def_QCD}) may be integrated out, and this leads to a
contribution $\sum\log(\det(D\!\!\!\!/+\!m^{(i)}))$ to the effective
action, where the sum runs over the flavors, $i=1,...,\Nf$.
Todays lattice terminology attributes the label $\Nf=2$ to simulations with a
common mass $m_{ud}$ of the ``sea'' quarks (i.e.\ those which come from the
functional determinant).
Studies which include a dynamical strange and possibly a dynamical charm quark
are referred to as $\Nf=2+1$ and $\Nf=2+1+1$ simulations, respectively, to
indicate that these quarks have separate masses.

The lattice spacing and the quark masses cannot be dialed ``a priori'' because
of renormalization effects.
The simulations are governed by the bare gauge coupling $\be=2\Nc/g^2$ and
several (in the cases mentioned above: $2$ or $2+1$ or $2+1+1$) bare mass
parameters; the lattice spacing $a$ and the quark masses $m_q$ are emergent
quantities to be determined ``a posteriori''.

\clearpage

%%%%%%%%%%%%%%%%%%%%%%%%%%%%%%%%%%%%%%%%%%%%%%%%%%%%%%%%%%%%%%%%%%%%%%%%%%%%%%%

\section{Hadron Spectroscopy \label{sec:spectroscopy}}

%%%%%%%%%%%%%%%%%%%%%%%%%%%%%%%%%%%%%%%%%%%%%%%%%%%%%%%%%%%%%%%%%%%%%%%%%%%%%%%

\subsection{Quenched versus unquenched QCD}

Spectroscopy of stable states with a conserved quantum number (e.g.\ isospin)
is about the easiest thing to do in Lattice QCD.
One considers two-point functions
\beq
C(t)=\<O(t)O\dag(0)\>=\frac{1}{Z}\int\!DU\;
\underbrace{\hat{O}(t)\hat{O}\dag(0)}_\mr{``valence''}\;
\prod_{i=1}^{\Nf}\underbrace{\det(D\!\!\!\!\slash\,[U]\!+\!m^{(i)})}_\mr{``sea''}\;
e^{-S[U]}
\eeq
where $O(.)$ is designed to absorb these quantum numbers, and at least one of
the two involves some projection to a definite spatial momentum (e.g.\
$\mb{p}=0$).
From the asymptotic behavior $C(t)\propto e^{-E(\mb{p})t}$ one extracts the
mass of the particle.
In the past the functional determinant was omitted ($\Nf=0$), while $O$ would
still contain so-called valence quarks; this is the quenched approximation.
If a quark mass in the determinant is different from the mass of the same
flavor in the interpolating fields $O(.)$, i.e.\
$m_{ud}^\mr{sea}\neq m_{ud}^\mr{val}$, one says that the respective flavor is
partially quenched.

\subsection{Landscape of current $\Nf=2+1$ simulations}

\begin{figure}[b]
\centering
\vspace*{-1mm}
\includegraphics[width=0.9\textwidth]{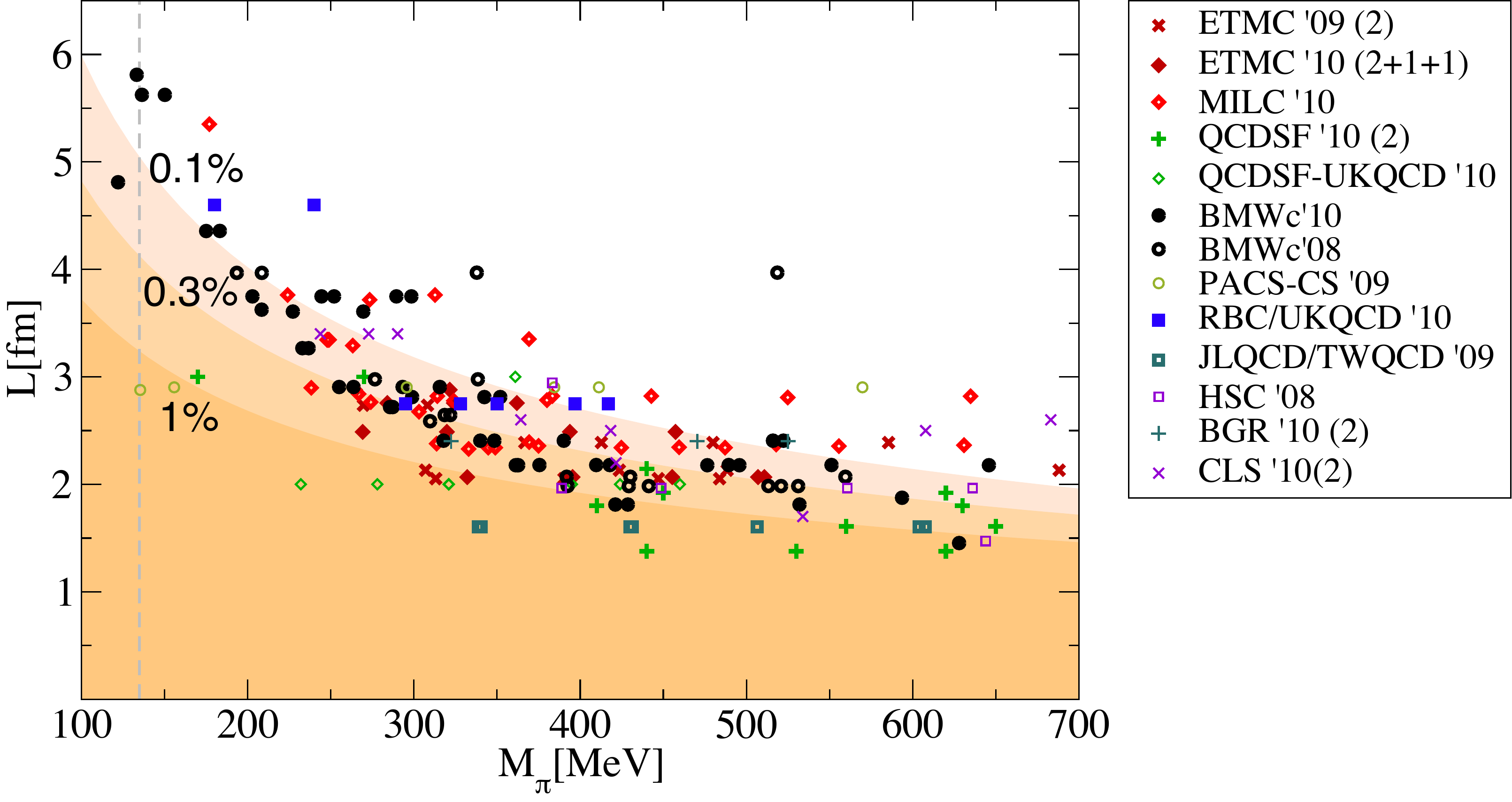}
\vspace*{-3mm}
\caption[*]{\label{fig:landscape}
Landscape of simulated box sizes and pion masses in recent studies of
full QCD. The amber shaded areas indicate the expected magnitude of relative
finite size effects.  Figure from \cite{Fodor:2012gf}.}
\end{figure}

As discussed before, generating ensembles with light $m_{ud}$ tends to be
expensive.
Fig.\,\ref{fig:landscape} depicts the sea pion masses and box sizes at which
various collaborations managed to simulate.
A significant number of ensembles is needed to extrapolate to
$(m_{ud}^\mr{phys},m_s^\mr{phys})$ and to remove the cut-off and finite-size
effects.
To reach the physical point most collaborations fix $m_s\simeq m_s^\mr{phys}$
and reduce $m_{ud}$ as much as possible (as of this writing only two
collaborations can bracket $m_{ud}^\mr{phys}$ by the $m_{ud}$ in the sea
\cite{Durr:2010aw,Bazavov:2012uw}).
By contrast, QCDSF pursues an interesting alternative.
They start with a common sea quark mass near
$(2m_{ud}^\mr{phys}\!+\!m_s^\mr{phys})/3$ and split the masses symmetrically
such that this weighted sum stays invariant \cite{Bietenholz:2011qq}.

\subsection{Reality check: spectra of stable hadrons}\vspace*{-2pt}

\begin{figure}[b]
\vspace*{-3mm}
\includegraphics[height=4.5cm]{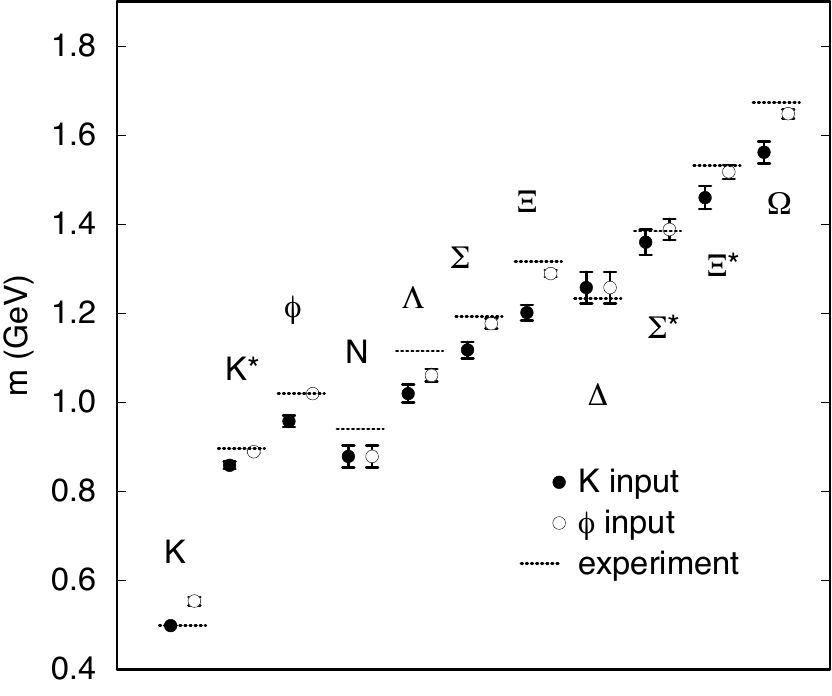}
\includegraphics[height=4.6cm]{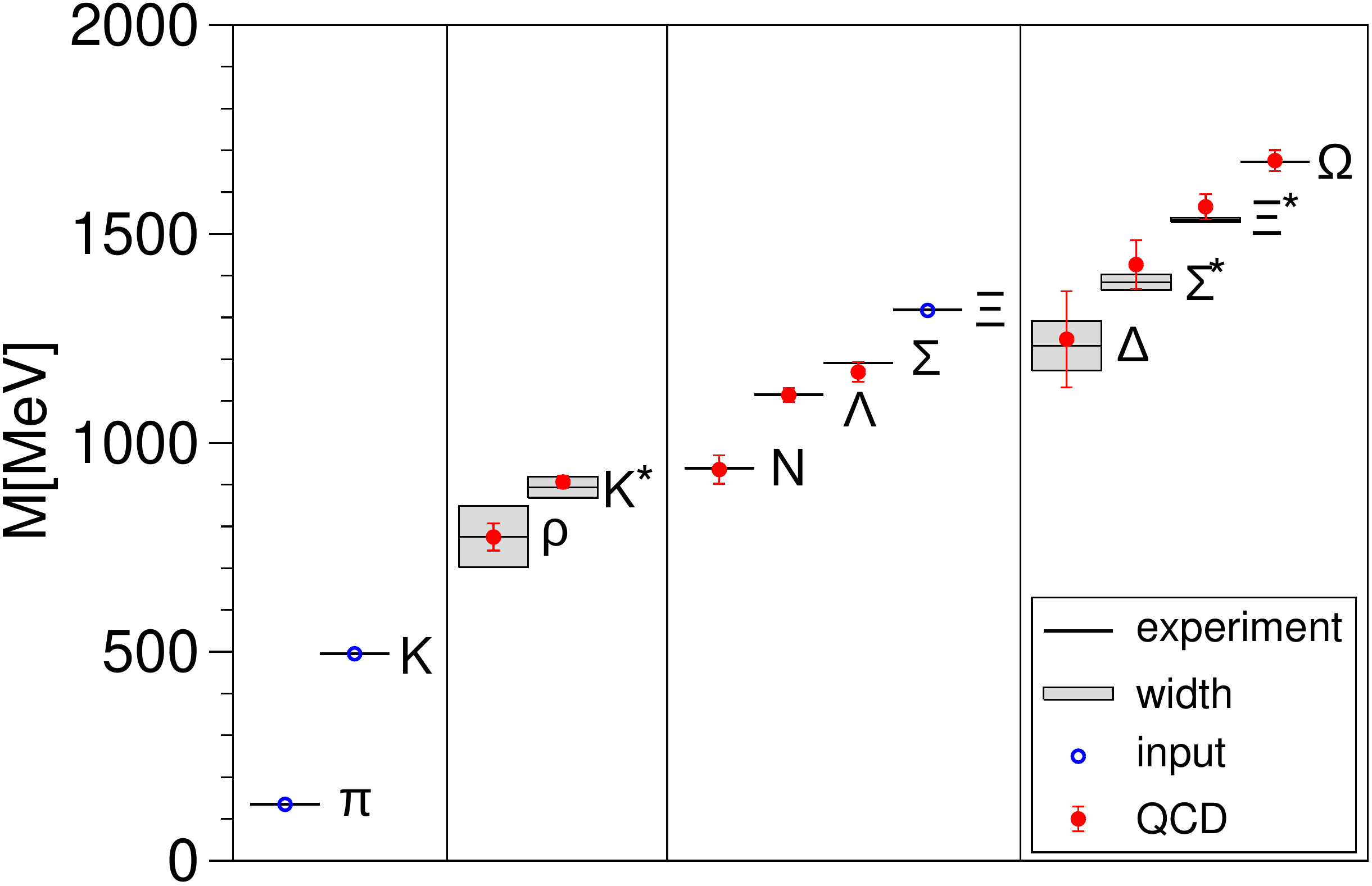}
\vspace*{-7mm}
\caption[*]{\label{fig:spectra}
Light pseudoscalar and vector meson spectra after $a\to0$, $L\to\infty$ in
quenched QCD (left, $\Nf=0$, published in 2000 \cite{Aoki:1999yr}) and in
full QCD (right, $\Nf=2+1$, published in 2008 \cite{Durr:2008zz}).}
\end{figure}

We tend to postulate that the QCD Lagrangian (\ref{def_QCD}) is the complete
theory of strong interactions, valid both in the short-distance regime (where
asymptotic freedom prevails) and in the long-distance regime (where confinement
dominates).

This is a highly non-trivial statement, and it was a historical achievement of
the CP-PACS collaboration to show that quenched QCD (as an alternative
candidate) does not pass the test \cite{Aoki:1999yr} (left panel in
Fig.\,\ref{fig:spectra}\,).
The unquenched version fares much better \cite{Durr:2008zz} ($\Nf=2+1$, right
panel in Fig.\,\ref{fig:spectra}\,); of course, this does not amount to a
proof.
In the latter case three quantities are plotted with open circles, as they have
been used to adjust $m_{ud}$, $m_s$ and to determine the individual $a$
(cf.\ Sec.\,\ref{sec:lattice}\,).

\subsection{Resonances and excited baryon spectra}\vspace*{-2pt}

Those who wish to tackle harder problems may study how resonances interact with
stable particles (note that the lattice results in Fig.\,\ref{fig:spectra}
concern only their masses; the gray bands indicate experimental widths).
A prominent example is $g_{\rh\pi\pi}$, the coupling of the rho to two pions,
on which there are nice results; see \cite{Mohler:2012nh} for a review.
In the same category is the study of excited states with more mundane quantum
numbers (e.g.\ those of the proton), see e.g.\ \cite{Bulava:2010yg} for recent
progress.

\subsection{Mixing of $\eta-\eta'$}\vspace*{-2pt}

One of the long standing problems of QCD is to convincingly show that the mass
splitting between the $\et$ and $\et'$ is due to the global axial anomaly (as
argued by Witten and Veneziano long ago).
This is now achieved on a quantitative level (i.e.\ with mixing) in two recent
papers, one by RBC/UKQCD \cite{Christ:2010dd}, one by ETMC
\cite{Ottnad:2012fv}.

\subsection{Progress on glueballs}\vspace*{-2pt}

Glueballs are hard in pure Yang-Mills theory (because they are noisy, and
sometimes a vacuum contribution needs to be subtracted), and they are extremely
hard in the presence of dynamical flavors (since they mix with other flavor
singlet states which do have valence quarks).
For a recent paper on the subject see \cite{Gregory:2012hu}.

\clearpage

%%%%%%%%%%%%%%%%%%%%%%%%%%%%%%%%%%%%%%%%%%%%%%%%%%%%%%%%%%%%%%%%%%%%%%%%%%%%%%%

\section{Hadron Structure \label{sec:structure}}

%%%%%%%%%%%%%%%%%%%%%%%%%%%%%%%%%%%%%%%%%%%%%%%%%%%%%%%%%%%%%%%%%%%%%%%%%%%%%%%

\subsection{Nucleon sigma terms and dark matter searches}\vspace*{-1pt}

\begin{figure}[b]
\vspace*{-7mm}
\includegraphics[height=4.4cm]{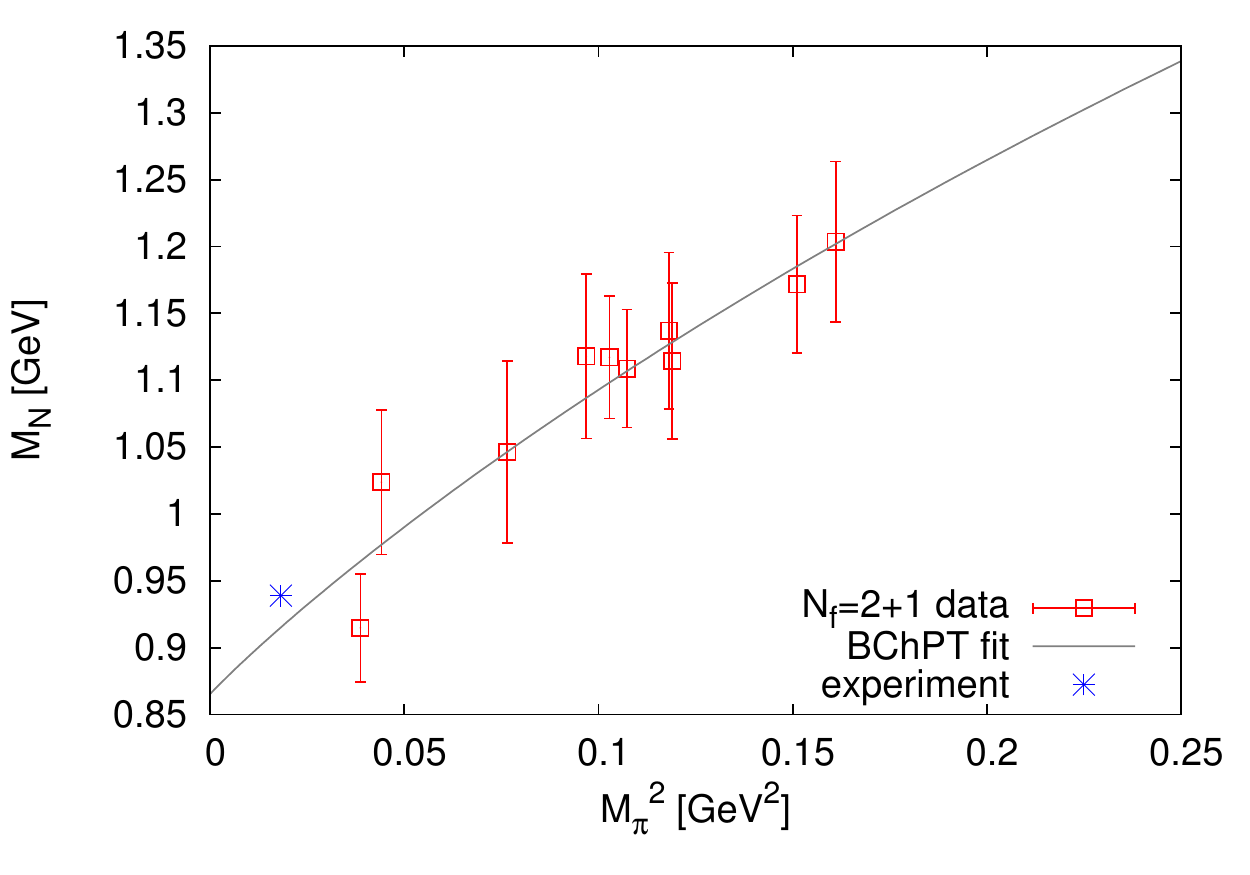}
\includegraphics[height=4.4cm]{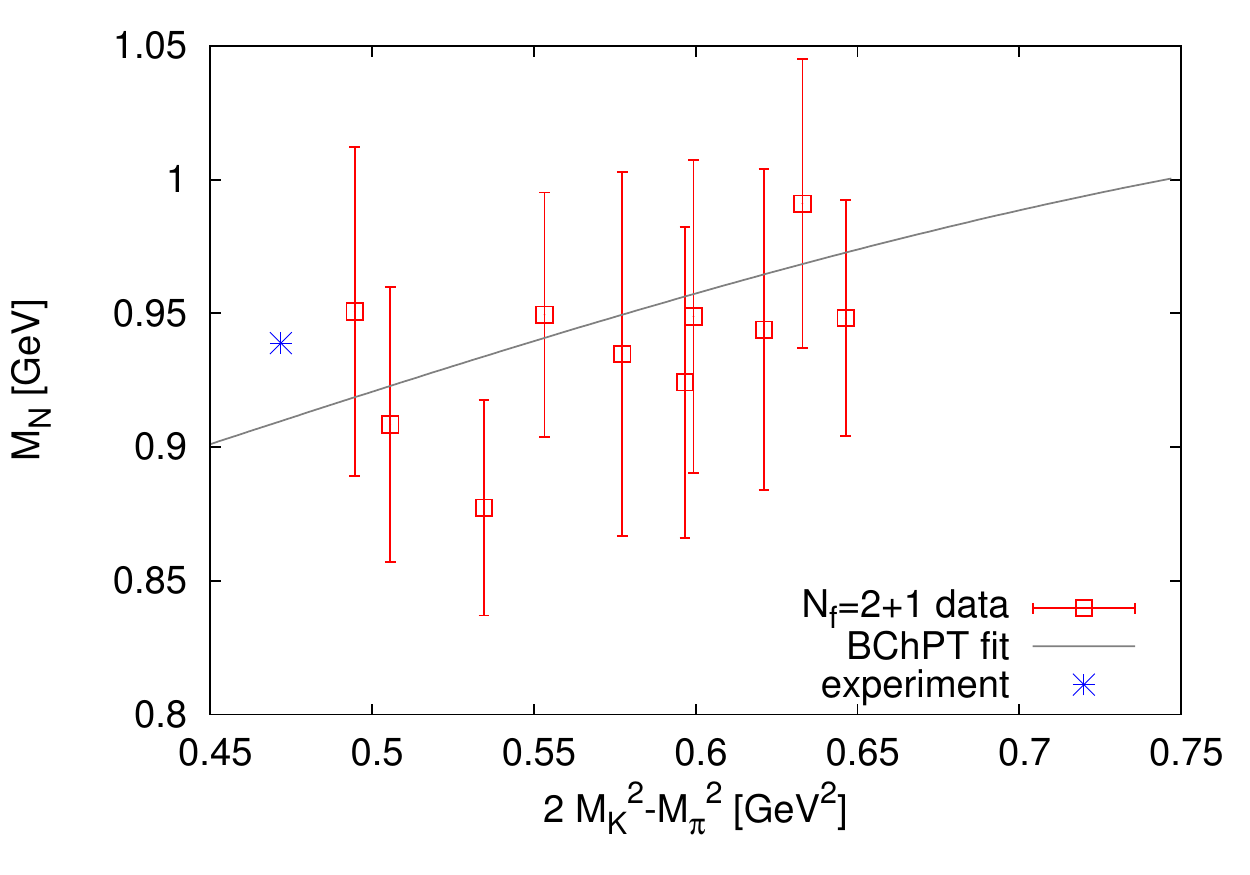}
\vspace*{-9mm}
\caption[*]{\label{fig:sigmaterms}
The nucleon mass as a function of the average light quark mass
$m_{ud}\propto\Mpi^2$ at fixed physical $m_s\propto2\Mka^2\!-\!\Mpi^2$ (left)
and vice versa (right) in full QCD simulations \cite{Durr:2011mp}.}
\end{figure}

Dark matter experiments try to probe the interaction of unknown objects
(WIMPs) with nuclei.
This would happen via coupling to virtual light and strange quark loops
in the nucleon, and the lattice is thus called to calculate the
quantities
\beq
\si_{\pi N}=\si_{ud}=m_{ud}\<N|u\bar{u}\!+\!d\bar{d}|N\>
\;,\qquad
\si_s=m_s\<N|s\bar{s}|N\>  %y_N=2\<N|s\bar{s}|N\>/\<N|u\bar{u}\!+\!d\bar{d}|N\>
\eeq
from first principles.
This can be done either from the slope of $M_N$ versus $\Mpi^2$ and
$2\Mka^2\!-\!\Mpi^2$ (see Fig.\,\ref{fig:sigmaterms}\,) or from a direct
determination of the matrix elements.
A straight average of all central values and total errors in
Tab.\,\ref{tab:sigmaterms} would suggest that
\beq
\si_{\pi N}=\si_{ud}=47(9)\MeV
\;,\qquad
\si_s=48(25)\MeV
\eeq
is a conservative estimate (in some cases $\si_s$ is obtained from
$y_N=2m_l/m_s\cdot\si_s/\si_{ud}$ or $f_{T_s}=m_s\<N|s\bar{s}|N\>/M_N$),
see also \cite{Young}.
With $\si_{\pi N}\simeq\Mpi^2\,\pad(M_N)/\pad(\Mpi^2)$ it follows
that the nucleon mass in the 2-flavor chiral limit is
$M_N(m_{ud}\!=\!0)\simeq880(20)\MeV$.

%%% datb(01,:)=[50,sqrt(9^2+3^2),33,sqrt(16^2+5^2)];
%%% datb(02,:)=[75,15,nan,nan];
%%% datb(03,:)=[nan,nan,59,sqrt(6^2+8^2)];
%%% datb(04,:)= nan* [59,sqrt(2^2+17^2),-4,sqrt(23^2+25^2)];
%%% datb(05,:)=[39,sqrt(4^2+12.5^2),34,sqrt(14^2+26^2)];
%%% datb(06,:)=[31,sqrt(3^2+4^2),71,sqrt(34^2+59^2)];
%%% datb(07,:)=[nan,nan,40,sqrt(7^2+5^2)];
%%% datb(08,:)=[45,6,21,6];
%%% datb(09,:)=[nan,nan,8,sqrt(14^2+15^2)];
%%% datb(10,:)=[43,sqrt(1^2+6^2),126,sqrt(24^2+54^2)];
%%% datb(11,:)=[nan,nan,43,10];
%%% datb(04,:)=[nan,nan,49,sqrt(10^2+15^2)]; % note: overwrite
%%% mean(datb(isfinite(datb(:,1)),1:2))
%%% mean(datb(isfinite(datb(:,3)),3:4))

\begin{table}[b]
\vspace*{-3mm}
\centering
\small
\begin{tabular}{|ccl|} %{|@{\,\,}c@{\,\,}c@{\,\,}l@{\,\,}|}
\hline
%%%
% $66.7(1.3)(?)$        & ------ & ETMC 08 \cite{Alexandrou:2008tn} \\ % si_l=66.7(1.3)(inf)
% $53(2)({+21\atop-7})$ & $22(12)({+10\atop-7})$ & JLQCD 08 \cite{Ohki:2008ff} \\ % si_l=53(2)(+21/-7) y_N=0.030(16)(+6/-8)
% $38(9)(8)$            & ------ & QCDSF 11 \cite{Bali:2011ks} \\ % si_l=38(12) f_T_s^{heavy}=0.012(14)^{+10}_{-3}
% $37(8)(6)$            & ------ & QCDSF 12 \cite{Bali:2012qs} \\ % si_l=37(8)(6)
% \hline
  $50(9)(3)$   & $33(16)(5)$   & Young\,Thomas 09 \cite{Young:2009zb}\\ % si_l=50(9)(1)(3) si_s=33(16)(4)(2)
  $75(15)$     & ------        & PACS-CS 09 \cite{Ishikawa:2009vc} \\ % si_l=75(15)
  ------       & $59(6)(8)$    & Toussaint\,Freeman 09 \cite{Toussaint:2009pz} \\ % si_s=59(6)(8)
% $59(2)(17)$  & $-4(23)(25)$  & Martin-Camalich et al 10 \cite{MartinCamalich:2010fp} \\ % si_l=59(2)(17) si_s=-4(23)(25)
  $39(4)({+18\atop-7})$ & $34(14)({+28\atop-24})$ & BMW-c 10 \cite{Durr:2011mp} \\ % si_l=39(4)(+18/-7) si_s=39(14)(+28/-24) y_N=0.20(7)(+13/-17)
  $31(3)(4)$   & $71(34)(59)$  & QCDSF/UKQCD 11 \cite{Horsley:2011wr} \\ % si_l=31(3)(4) si_s=71(34)(59)
  ------       & $40(7)(5)$    & MILC 12 \cite{Freeman:2012ry} \\ % <N|sbars|N>=0.44(8)(5) in (MSbar,2GeV) [use m_s=90MeV]
  $45(6)$      & $21(6)$       & Shanahan et al 12 \cite{Shanahan:2012wh} \\ % si_l=45(6) si_s=21(6)
  ------       & $8(14)(15)$   & JLQCD 12 \cite{Oksuzian:2012rzb} \\ % f_Ts=0.009(15)(16)
  $43(1)(6)$   & $126(24)(54)$ & Ren et al 12 \cite{Ren:2012aj} \\ %si_l=43(1)(6), si_s=126(24)(54)
  ------       & $43(10)$      & Engelhard 12 \cite{Engelhardt:2012gd} \\ %f_Ts =0.046(11)
  ------       & $49(10)(15)$  & Junnarkar Walker-Loud 13 \cite{Junnarkar:2013ac} \\
%%%
% sil=[53,02,21,07]; y=[0.030,0.016,0.006,0.008]; sis=27.5/2*sil(1)*y(1); sis=sis*[1,sqrt((sil(2:4)/sil(1)).^2+(y(2:4)/y(1)).^2)]
\hline
\end{tabular}
\vspace*{-2mm}
\caption[*]{\label{tab:sigmaterms}
Summary of recent $\Nf=2+1$ lattice determinations of $\si_{\pi N}$ and/or
$\si_s$.}
\end{table}

\subsection{Nuclear structure and recent news on $g_A$}\vspace*{-1pt}

The form factors of the nucleon with external $S,P,V,A$ currents and the
respective radii are known to be hard on the lattice.
Even the values at $\mb{p}^2\!=\!0$, in case of the axial current known as
$g_A$, turn out to be unexpectedly hard \cite{Brandt:2011sj,Green:2012ud}
(see also \cite{Lin}).

\subsection{Scattering of $\pi\pi$, $\pi K$, $KK$, $\pi N$, $NN$ and more}

\begin{figure}[b]
\vspace*{-8mm}
\begin{minipage}{6.5cm}
\includegraphics[height=5.0cm]{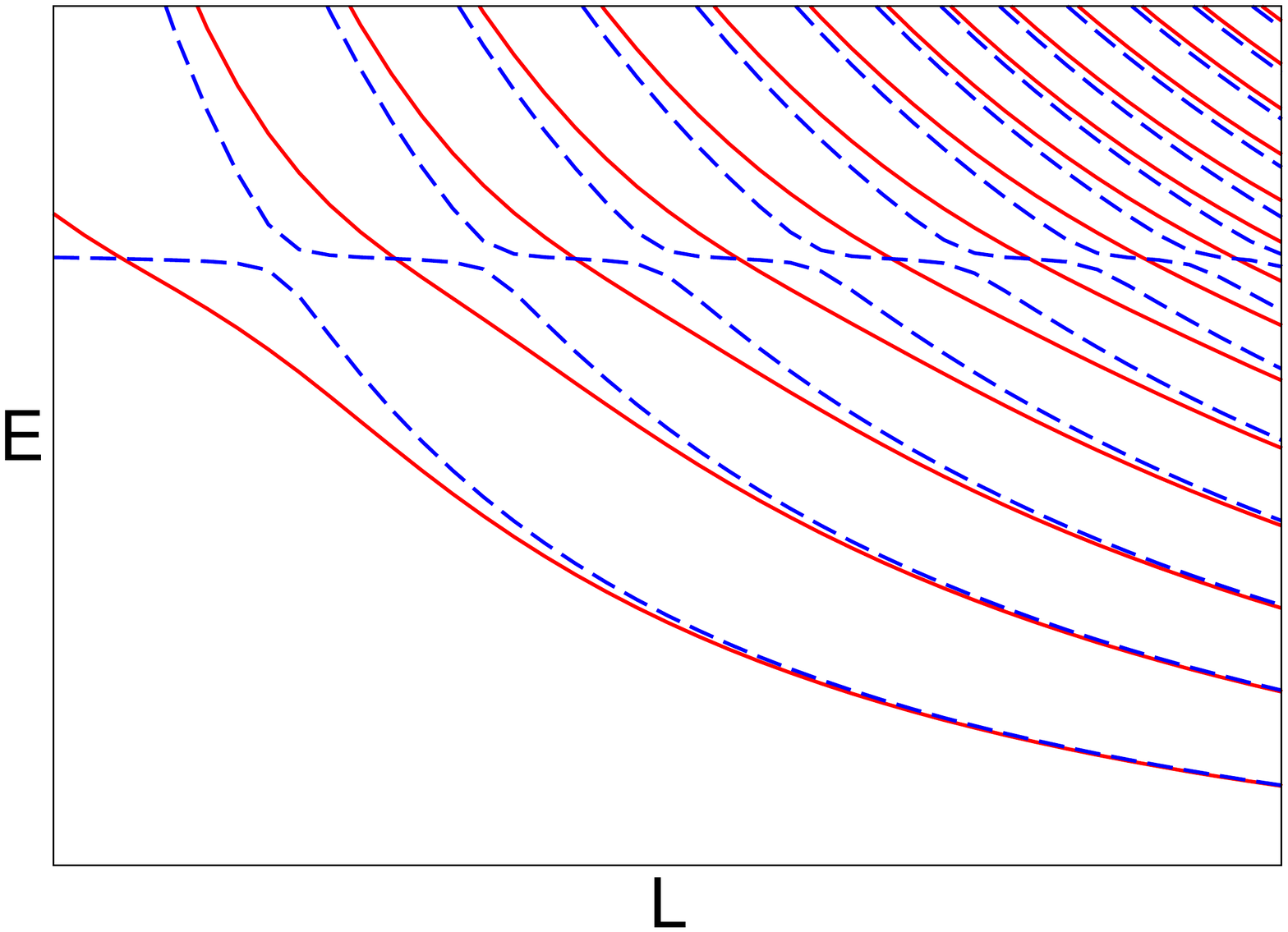}
\end{minipage}
\begin{minipage}{6.0cm}
\includegraphics[height=5.8cm,angle=-90]{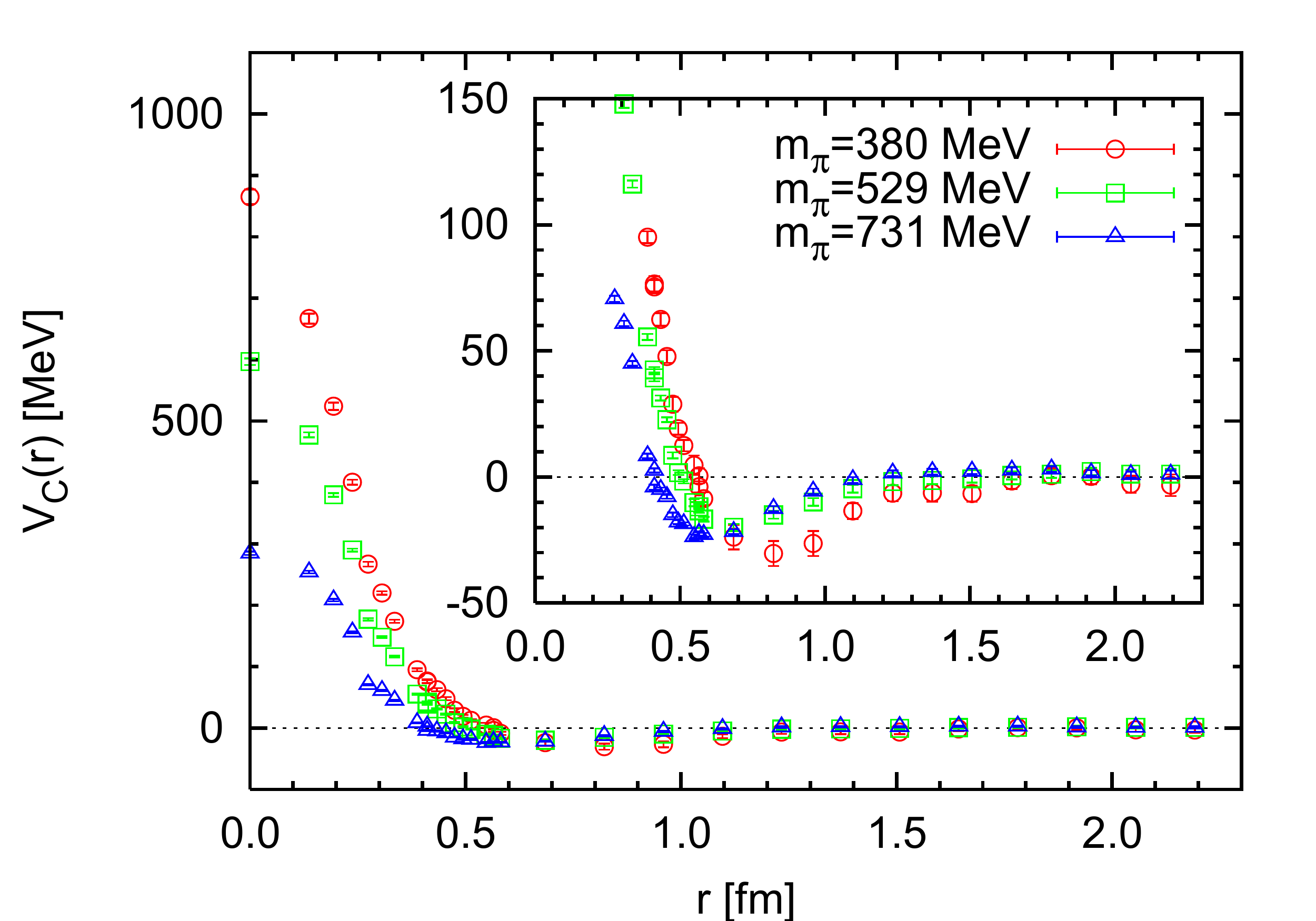}
\end{minipage}
\vspace*{-7mm}
\caption[*]{\label{fig:potential}
Principle of extracting scattering lengths and phase shifts from
avoided level crossings in a finite volume (left, ``level method''
\cite{Luscher:1990ck}, figure from \cite{Fodor:2012gf}) and result for central
nucleon-nucleon potential in ${}^1S_0$ state at various pion masses (right,
``potential method'', figure from \cite{Aoki:2009ji}).}
\end{figure}

Elastic scattering parameters can be extracted from the $L$-dependence of
accurately measured energies of two-body states in a finite volume
\cite{Luscher:1990ck}, cf.\ Fig.\,\ref{fig:potential} left.
Recently a new method has been proposed to extract such information from
two-body potentials, and it is claimed that such results are less sensitive
to excited states contaminations \cite{Aoki:2009ji}.
The right panel shows the central $NN$ potential (spin-singlet channel)
determined with this method; both the repulsion at short distance and the
attraction at intermediate distance seem to grow as the pion mass decreases.

\subsection{From quarks to nuclei}

\begin{figure}[b]
\centering
\vspace*{-5mm}
\includegraphics[width=0.77\textwidth]{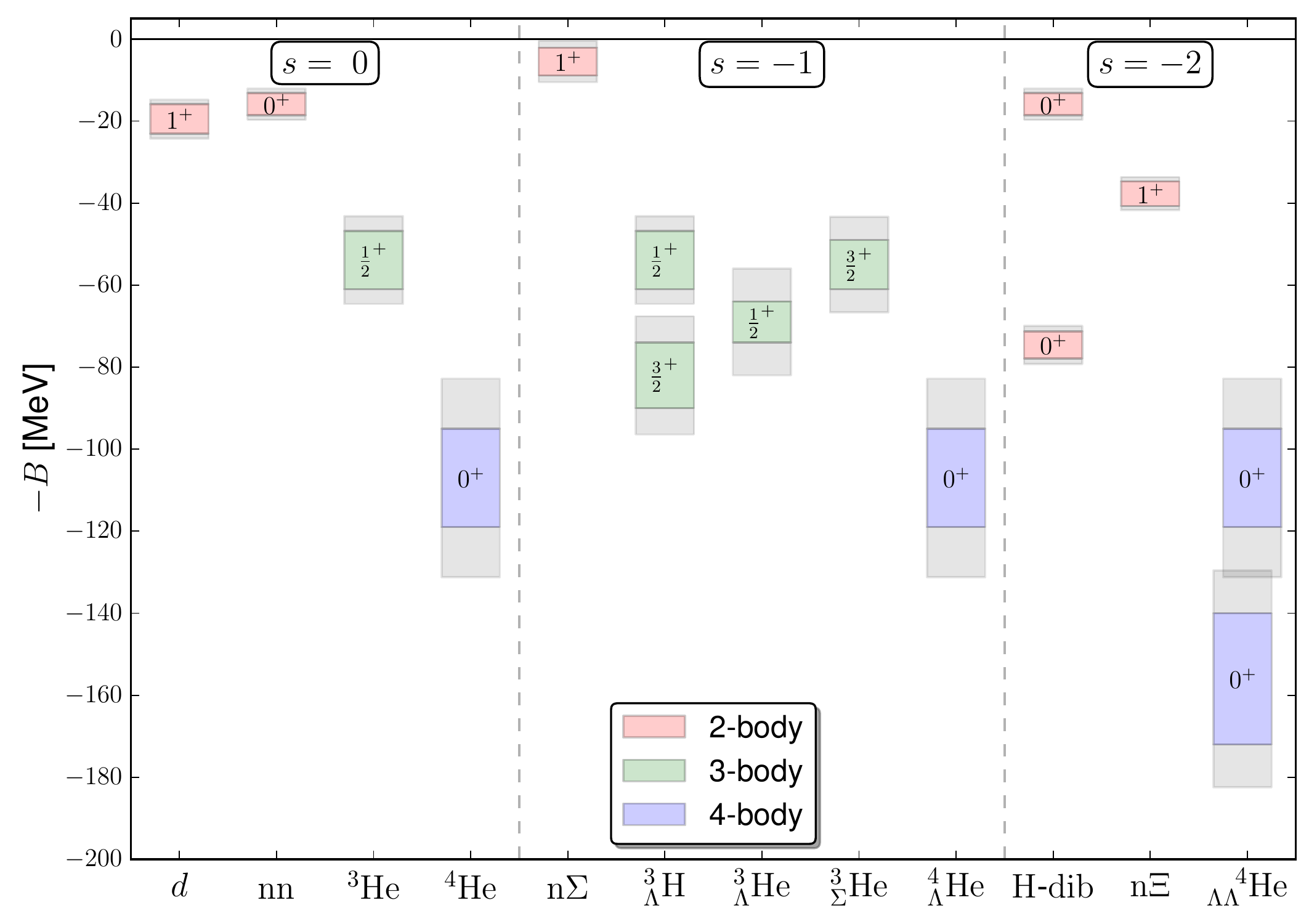}
\vspace*{-4mm}
\caption[*]{\label{fig:nplqcd}
A compilation of the nuclear energy levels, with spin and parity, as determined
in \cite{Beane:2012vq}\,.}
\end{figure}

Evidently, there is a long way to go until all nuclear physics can be derived
from Lattice QCD; see Fig.\,\ref{fig:nplqcd} and
\cite{Beane:2012vq,Yamazaki:2012hi}.
A method to deal with the vastly growing number of contractions is proposed in
\cite{Detmold:2012eu}.
The subject is reviewed in \cite{Doi:2012ab}.

\clearpage

%%%%%%%%%%%%%%%%%%%%%%%%%%%%%%%%%%%%%%%%%%%%%%%%%%%%%%%%%%%%%%%%%%%%%%%%%%%%%%%

\section{Flavor Physics and FLAG \label{sec:flavor}}

%%%%%%%%%%%%%%%%%%%%%%%%%%%%%%%%%%%%%%%%%%%%%%%%%%%%%%%%%%%%%%%%%%%%%%%%%%%%%%%

\subsection{Light quark masses and $\al_\mr{strong}$}

The quark masses and $\al_\mr{strong}$ are the parameters of the Standard
Model that Lattice QCD can determine without further theoretical input.
The starting point is a spectroscopy calculation like the one shown in the
right panel of Fig.\,\ref{fig:spectra}\,.
With this one has the quark masses in hand, albeit in an awkward scheme which
is specific to the lattice action used in the course of the simulation.
Before taking the continuum limit a conversion to a continuum scheme
(e.g.\ $\MSbar$ or $\mr{RI/MOM}$ or $\mr{SF}$) must be performed.
A few recent computations are listed in Tab.\,\ref{tab:qmasses}\,.

\begin{table}[b]
\vspace*{-2mm}
\centering
\small
\begin{tabular}{|ccccl|}
\hline
88(0)(5)       & 3.2(0)(2)     & 1.9(0)(2)    & 4.6(0)(3)    & MILC \cite{Bazavov:2009bb} \\
92.2(1.3)      & ------        & 2.01(10)     & 4.77(15)     & HPQCD \cite{McNeile:2010ji} \\
95.5(1.1)(1.5) & 3.469(47)(48) & 2.15(03)(10) & 4.79(07)(12) & BMW-c \cite{Durr:2010vn} \\
94.2(1.4)(5.7) & 3.31(07)(26)  & 1.90(08)(23) & 4.73(09)(36) & LVdW \cite{Laiho:2011np} \\
92.3(1.9)(1.3) & 3.37(09)(07)  & ------       & ------       & RBC/UKQCD \cite{Arthur:2012yc} \\
\hline
\end{tabular}
\vspace*{-2mm}
\caption[*]{\label{tab:qmasses}
Selection of recent computations of $m_s,m_{ud},m_u,m_d\,[\mr{MeV}]$ in the
$(\MSbar,2\GeV)$ scheme.}
\end{table}

Similarly, by observing how the lattice spacing $a$ varies as one shifts the
coupling $\be=2\Nc/g^2$ or by matching small Wilson loops to lattice
perturbation theory one can determine the QCD $\be$-function and the
integration constant $\Lambda_\mr{QCD}$ (the case $\Nf=2$ has been presented at
this conference \cite{BjornLeder}) or $\al_\mr{strong}(\mu)$ at a scale
$\mu\!\sim\!a^{-1}$.
The PDG average of $\al_\mr{strong}(M_Z)$ is dominated by the HPQCD results
\cite{Davies:2008sw,McNeile:2010ji}.

\subsection{Light decay constants and form factors}

The leptonic decay of a pseudoscalar meson ($P=\pi,K,...$) is mimicked on the
lattice by the coupling to an external axial current and captured in the
decay constant $f_P$.
Experiment determines the width $\Gamma\propto|f_\pi V_{ud}|^2$ or 
$\Gamma\propto|f_K V_{us}|^2$, and by dividing out $f_P$ one gets access to
the CKM matrix element $V_{ud}$ or $V_{us}$, respectively.

Similarly, the lattice can determine the $K\to\pi$ transition form factor
(with a vector current in between) $f_{+,0}^{K\to\pi}(q^2)$, and by combining
it with the width of the semileptonic $K\to\pi$ decay one has another option
to access $V_{us}$.

In either case, when computing $|V_{ud}|^2+|V_{us}|^2+|V_{ub}|^2$ (the
last contribution is tiny) one finds that the first-row CKM unitarity
relation is well satisfied.

\subsection{Kaon mixing in and beyond the standard model}

The $K^0$-$\bar{K}^0$ oscillation is the main source of indirect CP violation.
In the standard model (SM) it is dominated by two box diagrams which, via an
effective field theory approach, may be shrunk into a $\Delta\!S=2$ operator,
such that one ends up with $\<\bar{K}^0|O_{VV+AA}|K^0\>$.
The lattice can calculate this matrix element (which
enters the unitarity triangle analysis) or $B_K$, see e.g.\
\cite{Durr:2011ap,Bae:2011ff,Laiho:2011np,Arthur:2012yc}, as well as siblings
which are relevant to beyond standard model (BSM) theories
\cite{Boyle:2012qb,Bertone:2012cu}.

Recently RBC/UKQCD managed to calculate the $K\to(2\pi)_{I=2}$ amplitude (both
Re and Im of this $\Delta\!I=3/2$ process) \cite{Blum:2011ng}.
Still, the $\Delta\!I=1/2$ counterpart $K\to(2\pi)_{I=0}$ (relevant to
the $\Delta\!I=1/2$ rule and $\ep'/\ep$) is significantly harder.

\subsection{FLAG compilation and lattice averages}

The quantities discussed in the previous subsections have been evaluated
by many lattice groups.
In such a situation it is useful to have reviews which provide averages, like
the one by FLAG \cite{Colangelo:2010et} (with a focus on low-energy physics)
or \cite{Laiho:2009eu} (with an eye on CKM physics).
In either work the user is urged to cite the original literature.
Recently these two teams merged into FLAG-II with the goal to provide a
regularly updated compilation of a larger number of observables, with
trustworthy assessment of the overall systematic uncertainty:
\url{http://itpwiki.unibe.ch/flag}\,.

\subsection{Charm physics on the lattice}

\begin{figure}[b]
\vspace*{-8mm}
\includegraphics[height=4.4cm]{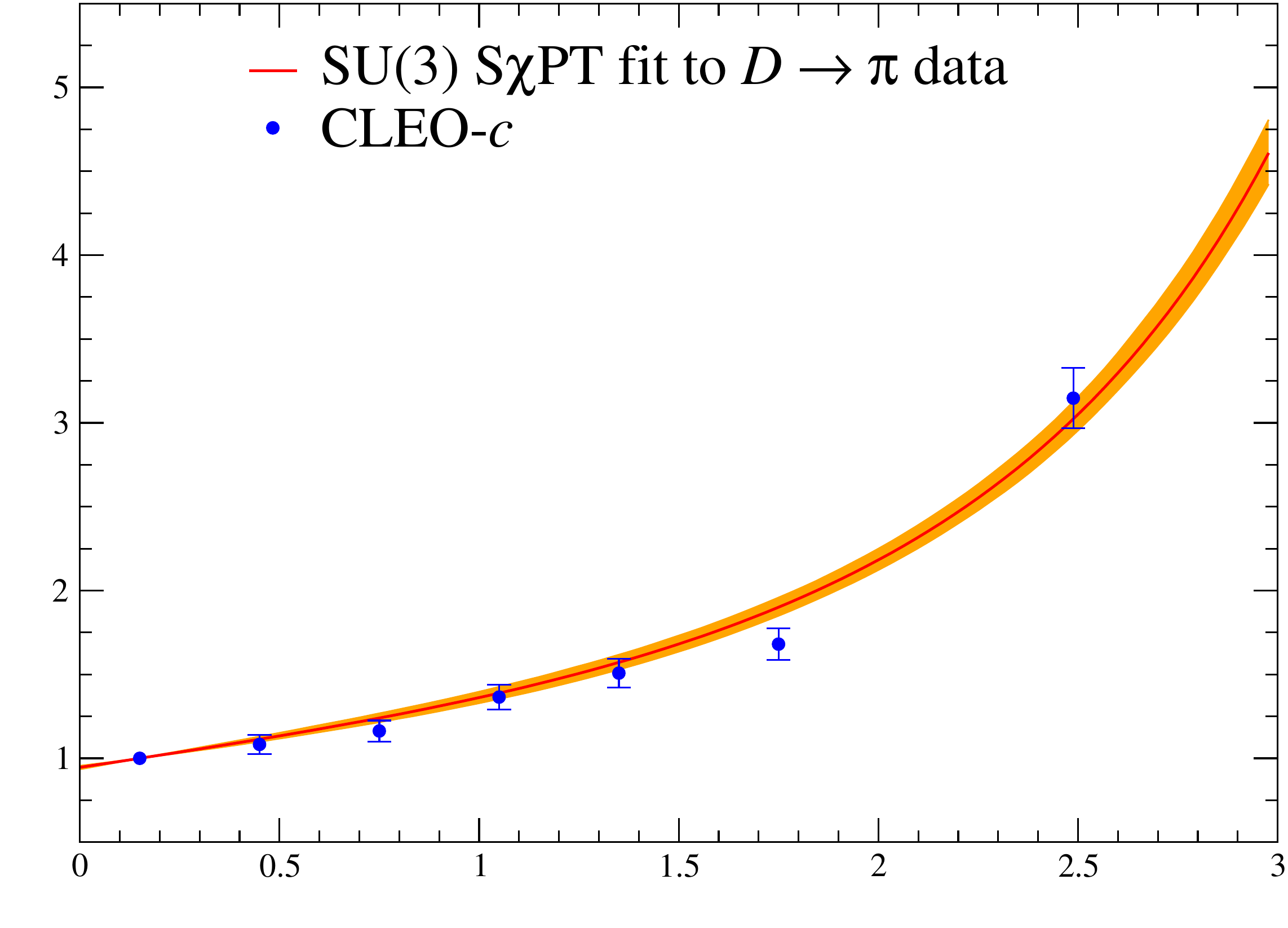}
\includegraphics[height=4.4cm]{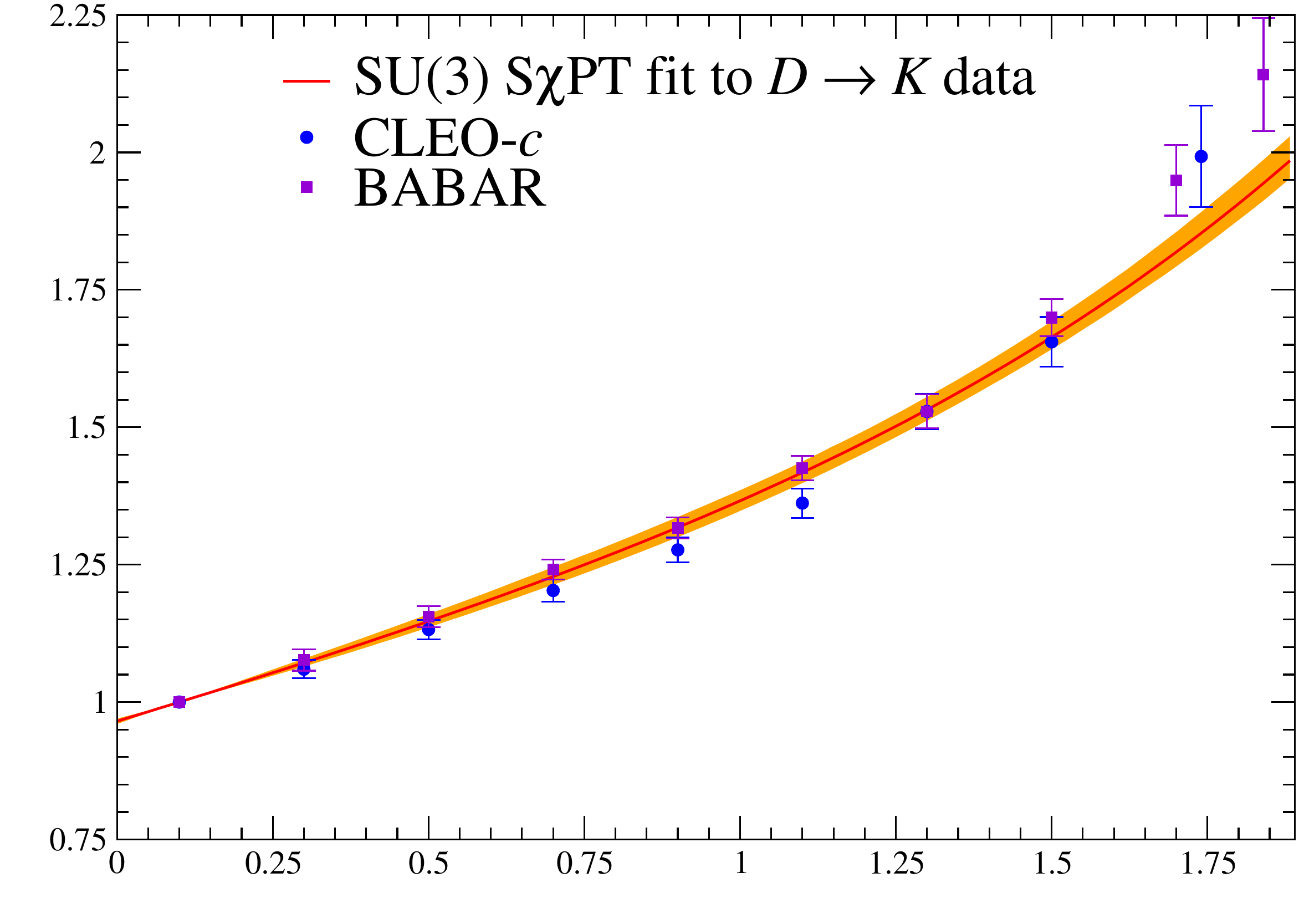}
\vspace*{-4mm}
\caption[*]{\label{fig:shape}
Shape of the form factors $f_+^{D\to\pi}(q^2)$ (normalized at $q^2=0.15\GeV^2$,
left) and $f_+^{D\to K}(q^2)$ (normalized at $q^2=0.10\GeV^2$, right) compared
to experiment. Figure from \cite{Bailey:2012sa}.}
\end{figure}

Treating the charm quark on the lattice just became doable
(cf.\ Sec.\,\ref{sec:lattice}).
Recent years brought several precise determinations of (leptonic) decay
constants and (semi-leptonic) transition form factors of $D$ and $D_s$ mesons
[essentially the same story as was told for pions and kaons before]; see
Fig.\,\ref{fig:shape} and Tab.\,\ref{tab:charm} for a tiny selection.

\begin{table}[b] %%% source 1210.8431
\vspace*{-6mm}
\centering
\begin{tabular}{|cccl|}
\hline
226(6)(1)(5)    & 257(2)(1)(5)    & 1.14(3)(0)(0) & PACS-CS \cite{Namekawa:2011wt} \\
212(8)          & 248(6)          & 1.17(5)       & ETMC \cite{Dimopoulos:2011gx} \\
208.3(1.0)(3.3) & 246.0(0.7)(3.5) & 1.187(04)(12) & HPQCD \cite{Na:2012iu} \\
209.2(3.0)(3.6) & 246.4(0.5)(3.6) & 1.175(16)(11) & FNAL/MILC \cite{Bazavov:2012dg} \\
\hline
\end{tabular}
\vspace*{-2mm}
\caption[*]{\label{tab:charm}
Selection of recent computations of $f_D\,[\mr{MeV}]$, $f_{D_s}\,[\mr{MeV}]$
and $f_{D_s}/f_D$.}
\end{table}

\subsection{Bottom physics on the lattice}

Due to its large mass the bottom quark tends to be treated with specialized
methods (cf.\ Sec.\,\ref{sec:lattice}).
Mass splittings in the $\Upsilon$ system and again decay constants and form
factors are of interest; see Tab.\,\ref{tab:bottom} for a spotlight and
\cite{Monahan:2012xx,Tarantino} for recent reviews.

\begin{table}[b]
\vspace*{-6mm}
\centering
\begin{tabular}{|cccl|}
\hline
195(12)    & 232(10)     & 1.19(5)   & ETMC \cite{Dimopoulos:2011gx} \\
------     & 225(04)     & ------    & HPQCD \cite{McNeile:2011ng} \\
196.9(9.1) & 242.0(10.0) & 1.229(26) & FNAL/MILC \cite{Bazavov:2011aa} \\
191(9)     & 228(10)   & 1.188(18) & HPQCD \cite{Na:2012kp} \\
\hline
\end{tabular}
\vspace*{-2mm}
\caption[*]{\label{tab:bottom}
Selection of recent computations of $f_B\,[\mr{MeV}]$, $f_{B_s}\,[\mr{MeV}]$
and $f_{B_s}/f_B$.}
\end{table}

\clearpage

%%%%%%%%%%%%%%%%%%%%%%%%%%%%%%%%%%%%%%%%%%%%%%%%%%%%%%%%%%%%%%%%%%%%%%%%%%%%%%%

\section{Other Topics \label{sec:other}}

%%%%%%%%%%%%%%%%%%%%%%%%%%%%%%%%%%%%%%%%%%%%%%%%%%%%%%%%%%%%%%%%%%%%%%%%%%%%%%%

\subsection{Large $N_c$, larger $N_f$, different representations}\vspace{-1mm}

Having a computer code which generates full QCD ensembles, it is 
straightforward to change the number of colors, $\Nc$, the number of dynamical
fermions, $\Nf$, or even their color representation.
Most practitioners are interested in the regime where such theories are
slowly walking, i.e.\ near conformality, see
\cite{Giedt,Panero:2012qx} for reviews.

\subsection{QCD thermodynamics at $\mu=0$ and $\mu>0$}\vspace{-1mm}

At vanishing baryon number density ($\mu\!=\!0$) and physical quark masses QCD
shows a crossover \cite{Aoki:2006we,Aoki:2009sc}.
Recently, a consensus was reached on the pseudocritical temperatures $T_c$ for
a set of common observables (subtracted chiral condensate, Polyakov loop
susceptibility) at physical quark masses and in the continuum
\cite{Aoki:2009sc,Borsanyi:2010bp,Bazavov:2011nk}.

QCD thermodynamics at non-zero chemical potential $(\mu\!>\!0)$ is in a much
less satisfactory state.
The main reason is that there is a genuine sign problem in the regime
$\mu\gg T,\Mpi$ (without any solution in sight).
There are established techniques to investigate the regime $\mu\ll T,\Mpi$
(which is good enough for cosmology, since the strong transition was at
$\mu\simeq0$).
To date it is not clear whether there is a line of first-order transitions
separating the confined (hadronic) phase from the deconfined (plasma) phase.
If it exists, it would terminate in a second-order endpoint \cite{Lombardo}.

\subsection{Hadronic contributions to $g\!-\!2$ of the muon}\vspace{-1mm}

Hadronic contributions to vacuum polarization provide one of the major sources
of systematic uncertainty in the computation of the anomalous magnetic moment
of the muon $a_\mu\equiv(g_\mu\!-\!2)/2$.
The lattice can help by providing the (Fourier transformed) 2-point function of
the electromagnetic vector current in such a hadronic environment.
Recent papers include \cite{Feng:2011zk,Boyle:2011hu,DellaMorte:2011aa};
see also \cite{Blum}.

\subsection{QCD with isospin splitting and/or electromagnetism}\vspace{-1mm}

In current state-of-the art studies of lattice QCD photon loops are excluded
from the simulations, and strong isospin breaking (due to $m_u\neq m_d$) is
ignored (since it is of the same order of magnitude).
To compensate, input quantities (e.g.\ $\Mpi,\Mka,M_\Xi$ in
Fig.\,\ref{fig:spectra} right) are usually adjusted for these effects,
see \cite{Colangelo:2010et} for details.

Recently, there has been significant progress in exploratory studies which
include quenched photons (i.e.\ interacting with the valence but not with the
sea quarks).
The key issue is that the finite-volume effects are no longer
exponentially small for large $\Mpi L$, but only polynomially suppressed
%e.g.\ $\Delta\!M^2(\infty)\!-\!\Delta\!M^2(L)\propto 1/L^2$
(see e.g.\ \cite{Blum:2010ym}).
Similarly, there has been tremendous progress in splitting $m_u\!\neq\!m_d$
through reweighting techniques (see \cite{Aoki:2012st}, though in this work a
considerable fraction of the reweighting power goes into shifting the average
light quark mass downwards).

%\subsection{Outlook: $N_f=1\!+\!1\!+\!1\!+\!1$ plus QED simulations}\vspace{-1mm}

The ultimate dream of todays lattice physicists is to perform
$N_f=1\!+\!1\!+\!1\!+\!1$ simulations, such that the physical values of
$m_u$, $m_d$, $m_s$, $m_c$ are bracketed by those in the ensembles, and the
lattices would include the photon as a dynamical degree of freedom and cover,
furthermore, several lattice spacings $a$ and box sizes $L$ such that an
extrapolation $a\to0$ and $L\to\infty$ can be performed.
%Given the progress sketched in the previous subsection, it seems conceivable
%that such studies can be performed in the not-so-distant future.

\subsection{Towards exaflop machines}\vspace{-1mm}

Among numerically oriented scientists of all disciplines the lattice
people are known for their hunger for computer time.
Current flagship machines tend to provide $O(10)$ petaflops of peak performance
[about $10^{16}$ multiply-and-add cycles per second], with a power
consumption of the order of 4\,MW.
Reaching the exaflop era ($10^{18}$ flops) requires new thoughts on how to make
these machines even more energy efficient, since scaling the electricity bill
up by two orders of magnitude is not an option.

The linpack tests that are used to rank such machines on the ``top\,500'' list
use about 80\% of the available cycles.
For typical lattice codes this sustained performance ratio ranges between
20\% and 50\% (which is significantly higher than what is reached in most other
disciplines).
Still, with every new generation of supercomputers the amount of parallelism
grows, and it is an ever increasing challenge to adjust the code such that
these high sustained performance figures would persist.

\subsection{More theoretical issues}\vspace{-1mm}

%From the outline in Sec.\,\ref{sec:lattice} it is clear that the lattice was
%originally designed for purely field-theoretic reasons \cite{Wilson:1974sk};
%the option of performing numerical studies came later\,\cite{Creutz:1980zw}.
%Even today, as large-scale numerical studies dominate the public perception,
%the lattice continues to be a key tool for progress in field theory.

Let me just mention some of the topics on which either tremendous progress
has been achieved or which mark open questions:
improved actions and matching with perturbation theory,
chiral symmetry in vector-like gauge theories,
chiral gauge theories and CP violation,
chiral symmetry and chemical potential,
the generic sign problem at non-zero chemical potential,
how one would formulate supersymmetry on the lattice,
further particulars of the staggered fourth-root procedure,
minimally doubled fermions and staggered fermions with non-standard mass terms,
the issue of large autocorrelation times near the chiral and/or continuum limit.

%%%%%%%%%%%%%%%%%%%%%%%%%%%%%%%%%%%%%%%%%%%%%%%%%%%%%%%%%%%%%%%%%%%%%%%%%%%%%%%

\section{Summary \label{sec:summary}}

%%%%%%%%%%%%%%%%%%%%%%%%%%%%%%%%%%%%%%%%%%%%%%%%%%%%%%%%%%%%%%%%%%%%%%%%%%%%%%%

Instead of giving a long summary, I try to formulate seven short messages:
\begin{enumerate}
\itemsep-1pt
\item
The discretization of the QCD Lagrangian (\ref{def_QCD}) [or something
equivalent] is a necessary intermediate step in the \emph{definition} of QCD.
\item
Spectroscopy of stable hadrons with $\Nf=2$ or $\Nf=2+1$ or $\Nf=2+1+1$
active (dynamical) quarks is a mature field.
\item
Spectroscopy of mixing or unstable [under strong interactions] states (often
performed on the same gauge backgrounds) is developing fast.
\item
Lattice QCD yields vital input in CKM analysis and BSM bounds, much of which is
reviewed in the compilations by FLAG \cite{Colangelo:2010et} and
latticeaverages \cite{Laiho:2009eu}.
Use these resources, but please be sure to cite the original papers !
\item
There is rapid progress on a number of nuclear issues, such as the nucleon
sigma term and various scattering lengths and phase-shifts.
\item
There is rapid progress on QCD thermodynamics. At $\mu=0$ a consensus on the
relevant pseudocritical temperatures $T_c$ has been reached, while at
$\mu\neq0$ it is still not clear whether a second-order endpoint exists.
\item
The lattice remains a relevant tool for conceptual work in field theory.
For instance, it is still not clear whether supersymmetry is a well-defined
field-theoretic concept (beyond perturbation theory).
\end{enumerate}

\bigskip
\noindent
{\bf Acknowledgements}:
The author wishes to thank the organizers of PIC'12 for enabling a very
enjoyable meeting in a truly superb place, as well as the other members of the
Budapest-Marseille-Wuppertal collaboration for a long standing cooperation.
The author is supported in part by the German SFB TRR-55.

%%%%%%%%%%%%%%%%%%%%%%%%%%%%%%%%%%%%%%%%%%%%%%%%%%%%%%%%%%%%%%%%%%%%%%%%%%%%%%%

%%%%%%%%%%%%%%%%%%%%%%%%%%%%%%%%%%%%%%%%%%%%%%%%%%%%%%%%%%%%%%%%%%%%%%%%%%%%%%%

\end{document}